# Jet Launching Structure Resolved Near the Supermassive Black Hole in M87

Authors: **Sheperd S. Doeleman**[1,3*], **Vincent L. Fish**[1], **David E. Schenck**[1,2†], Christopher Beaudoin[1], Ray Blundell[3], Geoffrey C. Bower[4], Avery E. Broderick[5,6], Richard Chamberlin[7], Robert Freund[2], Per Friberg[8], Mark A. Gurwell[3], Paul T. P. Ho[9], Mareki Honma[10,11], Makoto Inoue[9], Thomas P. Krichbaum[12], James Lamb[13], Abraham Loeb[3], Colin Lonsdale[1], Daniel P. Marrone[2], James M. Moran[3], Tomoaki Oyama[10], Richard Plambeck[4], Rurik A. Primiani[3], Alan E. E. Rogers[1], Daniel L. Smythe[1], Jason SooHoo[1], Peter Strittmatter[2], Remo P. J. Tilanus[8,14], Michael Titus[1], Jonathan Weintroub[3], Melvyn Wright[4], Ken H. Young[3], Lucy Ziurys[2]

**Affiliations:**

[1] MIT Haystack Observatory, Off Route 40, Westford, MA 01886, USA.

[2] Steward Observatory, Arizona Radio Observatory, University of Arizona, 933 N. Cherry Ave., Tucson AZ 85721-0065, USA.

[3] Harvard Smithsonian Center for Astrophysics, 60 Garden Street, Cambridge, MA 02138, USA.

[4] UC Berkeley, Department of Astronomy, Hearst Field Annex, Berkeley CA 94720 USA.

[5] Perimeter Institute, 31 Caroline St., N. Waterloo, Ontario, Canada N2L 2Y5.

[6] Dept. of Physics and Astronomy, University of Waterloo, 200 University Ave. W., Waterloo, On N2l 3G1, Canada.

[7] Caltech Submillimeter Observatory, 111 Nowelo St., Hilo, Hawai'i 96720, USA.

[8] James Clerk Maxwell Telescope, Joint Astronomy Centre, 660 N. A'ohoku Place University Park, Hilo, Hawaii 96720, U.S.A.

[9] Academia Sinica Institute for Astronomy and Astrophysics, 11F Astronomy-Mathematics Bldg., National Taiwan University, No. 1, Roosevelt Rd., Sec. 4 Taipei 10617, Taiwan R.O.C.

[10] National Astronomical Observatory of Japan, 2-21-1 Osawa, Mitaka, Tokyo 181-8588 Japan.

[11] The Graduate University for Advanced Studies, Osawa, Mitaka, Tokyo 181-8588, Japan.

[12] Max-Planck-Institut für Radioastronomie, Auf dem Hügel 69, 53121 Bonn, Germany.

[13] OVRO, California Institute of Technology, 100 Leighton Lane, Big Pine, CA 93513-0968 U.S.A.

[14] Netherlands Organization for Scientific Research, Laan van Nieuw Oost-Indie 300, NL2509 AC The Hague, The Netherlands.

[†] Current address: University of Colorado at Boulder, Dept. of Astrophysical and Planetary Sciences, 391 UCB, Boulder, CO, 80309 USA.

*To whom correspondence should be addressed. E-mail: sdoeleman@haystack.mit.edu

**Abstract**: Approximately 10% of active galactic nuclei exhibit relativistic jets, which are powered by accretion of matter onto super massive black holes. While the measured width profiles of such jets on large scales agree with theories of magnetic collimation, predicted structure on accretion disk scales at the jet launch point has not been detected. We report radio interferometry observations at 1.3mm wavelength of the elliptical galaxy M87 that spatially resolve the base of the jet in this source. The derived size of 5.5 +/- 0.4 Schwarzschild radii is significantly smaller than the innermost edge of a retrograde accretion disk, suggesting that the M87 jet is powered by an accretion disk in a prograde orbit around a spinning black hole.

**Main Text:** The compact central regions of some galaxies are so luminous that they outshine the combined output of all other energy sources in the galaxy. The small size and high power output of these active galactic nuclei (AGN) are most plausibly explained by the conversion of gravitational energy through accretion onto a super massive black hole. These black holes have masses of $M \gtrsim 10^6 \, M_\odot$, in contrast with lower mass black holes ($M \lesssim 10 \, M_\odot$) that result from the gravitational collapse of evolved stars. Many AGN produce powerful collimated jets of relativistic particles that can extend for hundreds and thousands of light-years, providing an important mechanism for redistributing matter and energy on large scales that affect galactic evolution (1). Jets are thought to form through magnetic acceleration processes located within the accretion flow or at the central black hole itself (2-4), but no observations to date have had the angular resolution required to detect and confirm structure on these scales for extragalactic jet sources. High-resolution radio interferometry of these sources at cm wavelengths is limited by optical depth effects that obscure the innermost accretion region. For these reasons, it remains unclear if jet formation requires a spinning black hole (5,6), and if so, whether jets are more likely to be formed when the orbital angular momentum of the accretion flow is parallel (prograde) or anti-parallel (retrograde) to the black hole spin (7,8). To address these questions, we have assembled a Very Long Baseline Interferometry (VLBI) array operating at a wavelength of 1.3mm, the Event Horizon Telescope (9), where AGN become optically thin, and angular resolutions necessary to resolve the inner accretion disks of nearby AGN are obtained.

Using a distance to M87 of 16.7 +/- 0.6 Mpc (10) and adopting the corresponding recent mass measurement of $(6.2 \pm 0.4) \times 10^9 \, M_\odot$ (11), the Schwarzschild radius of the M87 black hole ($R_{SCH} = 2GM/c^2 = (5.9 \pm 0.4) \times 10^{-4}$ parsec $= (1.9 \pm 0.12) \times 10^{15}$ cm) subtends an angle of $7.3 \pm 0.5$ micro arcseconds, presenting us with the best known opportunity for studying the formation of relativistic jets on scales commensurate with the black hole and accretion disk. Radiating via synchrotron emission, the relativistic jet from M87 extends for hundreds of kilo parsecs and terminates in extended lobes of emission as it slows and interacts with the intergalactic medium. Closer to the galaxy's core, on hundreds of parsec scales, the radio jet is remarkably well collimated with an opening angle of less than 5 degrees (12), and is also clearly seen in the optical, ultra-violet, and x-rays (13,14) where the emission is primarily confined to knots along the central 'spine' of the jet. VLBI observations at wavelengths ranging from 3.5mm to 20cm show that the jet opening angle, delineated by edge brightening in the outflow, continually increases as the core is approached, reaching ~60° within 1 milli arcsecond of the core (15-19). This wide opening angle is a signature of the launch point for a magnetohydrodynamically (MHD) powered jet that has not yet had time to collimate (2), and identifies the VLBI core as the most likely site of the central black hole.

We observed M87 over three consecutive days with a 1.3mm wavelength VLBI array consisting of four telescopes at three geographical locations: the James Clerk Maxwell Telescope (JCMT) on Mauna Kea in Hawaii, the Arizona Radio Observatory's Submillimeter Telescope (SMT) in Arizona, and two telescopes of the Combined Array for Research in Millimeter-wave Astronomy (CARMA, located ~60m apart) in California. On Mauna Kea, the JCMT partnered with the Submillimeter Array (SMA), which housed the Hydrogen maser atomic frequency standard and wideband VLBI recording systems; the SMT and CARMA were similarly equipped. These special-purpose systems allowed two frequency bands, each of 512 MHz bandwidth, to be sampled at 2-bit precision and recorded at an aggregate rate of 4 Gigabits/second. The two bands, labelled 'low' and 'high', were centered on 229.089 GHz and 229.601 GHz respectively. Data recorded at all sites were shipped to MIT Haystack Observatory for processing on the Mark4 VLBI correlator. Once correlated, data for each VLBI scan (typically 10 minutes) were corrected for coherence losses due to atmospheric turbulence and searched for detections using established algorithms tailored for high frequency observations (20). M87 was clearly detected each day on all VLBI baselines, and the interferometric data were then calibrated to flux density units (20).

Clear detections on the long baselines to Hawaii (CARMA-JCMT and SMT-JCMT) represent the highest angular resolution observations of M87 reported in any waveband, and when combined with the CARMA-SMT baseline data they provide a robust means to measure the size of the M87 core. The baseline between the two CARMA antennae corresponds to angular scales of ~4 arcseconds and is sensitive to extended and much larger scale jet structure; these data were thus used to refine the calibration of the antennas, but were excluded from analysis of the core component. To extract a size for the core, we fit a two-parameter circular Gaussian model to the 1.3mm VLBI data, deriving a total flux density and full width half maximum (FWHM) size for each day of observations. Sizes and flux densities fit separately for each day are consistent with each other at the $3\sigma$ level, indicating no significant variation in the 1.3mm core structure over the three days of observation (Fig. S4). When data from all three days are combined, the weighted least-squares best-fit model for the compact component results in a flux density of $0.98 \pm 0.04$ Jy and a FWHM of $40 \pm 1.8$ micro arcseconds ($3\sigma$ errors) (Fig. 1). Conversion to units of Schwarzschild radius yields a value of $5.5 \pm 0.4$ $R_{SCH}$ ($1\sigma$ errors) where the errors are dominated by uncertainties in the distance to M87 and the black hole mass. We adopt the circular Gaussian size derived using data from all three days for subsequent discussion.

Our VLBI observations cannot be used to fix the absolute position of this Gaussian component, however, in the case of M87 there is compelling evidence that this ultra-compact 1.3mm emission is in immediate proximity to the central super massive black hole. We first note that measurements of the jet width starting 10's of $R_{SCH}$ from the core and extending to core-separations of more than $10^4$ $R_{SCH}$ are well fit by a parabola-like collimating profile (19, Fig. 2). This fit matches similar profiles of General Relativistic MHD (GRMHD) jet simulations (21,22). When extrapolated to small scales, the empirical profile intersects the 1.3mm emission component size within one $R_{SCH}$ of the jet base. Because the angle of the M87 jet axis to our line of sight is estimated to be within the range $15^o – 25^o$ (23), the de-projected distance of this intersection point from the jet base lies in the range $2.5 – 4$ $R_{SCH}$. A second method of locating the 1.3mm emission derives from observations of the position shift of the M87 core as a function of wavelength. Multi-wavelength astrometric VLBI observations (15cm – 7mm) confirm that the absolute position of the core moves asymptotically towards the jet base with a $\nu^{-0.94}$ dependence (23). Extrapolation to 1.3mm wavelength places the 1.3mm VLBI component coincident with the jet base to within the uncertainty of the core-shift relation, which is ~1.5 $R_{SCH}$. In isolation, these observational trends would be unable to clearly link the jet base with the central engine. In blazar sources, for example, the relativistic jets are closely aligned to our line of sight and the jet base is illuminated hundreds of thousands of $R_{SCH}$ from the central engine (24). In contrast, M87 now represents a unique case where the jet base has a size (~5.5 $R_{SCH}$) that is consistent with scales on which energy is extracted from the black hole and accretion disk to feed the jet (2). It is thus most natural to spatially associate the 1.3mm VLBI component with the central engine, further guided by GRMHD simulations (21) that exhibit jet widths, within a few $R_{SCH}$ of the black hole, that match the 1.3mm VLBI size (Fig. 2). In M87, the favourable geometry of a misaligned jet and increased transparency of the synchrotron emission at mm wavelengths (25) has allowed us direct access to the innermost central engine with 1.3mm VLBI.

The most plausible mechanisms for powering extragalactic jets involve conversion of the black hole rotational energy through the Blandford-Znajek (BZ) process (3), whereby magnetic field lines cross the black hole event horizon and launch Poynting flux dominated outflows. The inner portion of the accretion disk is not only the source of the magnetic fields threading the black hole, but also launches a disk-wind via the Blandford-Payne (BP) mechanism (4), which serves to collimate the jet. This BZ/BP combination forms a spine/sheath morphology in which a narrow, electromagnetic and initially non-radiative jet from the black hole is surrounded by a slower and quickly mass-loaded outflow originating from the inner disk (6,26). In the case of M87, the broad opening angle and existence of a counter jet (17,18) indicate that the dominant contribution to the 1.3mm VLBI emission is from the slower moving sheath, anchored within the accretion disk (6). In this model of jet genesis, which we adopt here, the critical size scale associated with the jet footprint is the Innermost Stable Circular Orbit (ISCO) of the black hole, within which matter quickly plunges to the event horizon. The ISCO marks the peak density and rotational speed of the accretion flow (27), and is the location where particles are most efficiently accelerated from the disk (5) and begin to radiate. Strong beaming effects, which might produce small and high-brightness features unrelated to the ISCO, are not expected to be a factor on the scales probed by the 1.3mm VLBI observations because the detection of a counter jet in M87 and the parabolic profile of the jet both indicate that the outflow near the black hole is sub-relativistic.

Based on this understanding of the M87 jet, and taking the 1.3mm VLBI size as the ISCO diameter, we estimated the black hole spin and determined whether the accretion disk orbits in a prograde or retrograde sense. This is possible because the intrinsic ISCO diameter ($D_{ISCO}$) is sensitively dependent on black hole spin, with a value of $D_{ISCO} = 6\ R_{SCH}$ for a non-spinning (Schwarzschild) black hole ($a=0$), and ranging from $D_{ISCO} = 9\ R_{SCH}$ to $D_{ISCO} = 1\ R_{SCH}$ for retrograde and prograde orbits, respectively, around a maximally spinning (Kerr) black hole ($a=1$). Strong lensing effects due to the Kerr spacetime metric near the rotating black hole magnify the apparent size of the ISCO, with the relationship between observed ISCO size and black hole spin shown in Fig. 3 (28,20). The measured 1.3mm VLBI size corresponds to a prograde ISCO around a black hole with spin $a > 0.2$ (3σ). This result explicitly excludes the possibility of a retrograde ISCO orbit in the accretion flow because all such orbits would be larger than the core size derived here. The smallest possible retrograde ISCO orbit would present an apparent diameter of 7.35 $R_{SCH}$, more than 4σ larger than the observed size. This result is consistent with generally accepted theories that the spin axes of the accretion disk and black hole will be brought into alignment through gradual angular momentum transfer from the orbiting disk (29).

As the sensitivity and resolution of the 1.3mm VLBI array improves, modelling of the data can move beyond simple Gaussian distributions to physically motivated models that include accretion physics, relativistic beaming and full GR ray tracing. Recent modelling of the M87 jet on Schwarzschild radius scales indicates that in many cases emission from a counter-jet will illuminate the black hole from behind, creating a bright feature at the last photon orbit (25,30). The relatively dim region interior to this ring is known as the black hole shadow or silhouette, and its dimensions are determined by black hole mass, spin and inclination of the spin axis (31). Over a wide range of spin and inclination, the last photon ring of emission has a diameter of $D_{RING} = 5.2\ R_{SCH}$, so that fitting for the ring size yields an estimate of the black hole mass (32). The presence of such a shadow feature is not ruled out by the 1.3mm VLBI data presented here, which can be well fitted by a hybrid model combining a circular Gaussian representing jet emission launched from the ISCO, with a uniform annular ring at the last photon orbit (Fig. 1). Though the current VLBI array cannot be used to meaningfully constrain this more complex model, there are two distinguishing characteristics of this hybrid approach that can be readily tested with future 1.3mm VLBI arrays. The first is a predicted null in correlated flux density near baseline lengths of $4.5 \times 10^9\ \lambda$ ($\lambda$, observing wavelength), and the second is a 180-degree flip in interferometric phase between VLBI baselines on either side of this null.

It is increasingly clear that strong gravity effects can dominate observed AGN structure on the scales accessible with short wavelength VLBI. Included among these effects is the spin-dependent ISCO period, which ranges from 5 days ($a=1$) to 1 month ($a=0$) for the mass of the M87 black hole. The consistency of the 1.3mm VLBI sizes presented here, spanning three days of observation, does not reflect dramatic structural changes in the jet that might be expected due to accretion disk inhomogeneity for a black hole spinning near the maximum rate (Fig. S4). More sensitive searches for such periodic features in the jet launch region can be carried out with the full Event Horizon Telescope. In general, this work signals that Earth-sized 1.3mm VLBI networks are now able to provide angular resolutions that link observations of compact objects dominated by strong-field GR to outflows on the largest galactic scales.

**References and Notes:**

1. D. Richstone, *et al.*, Supermassive black holes and the evolution of galaxies, *Nature* **395**, A14-A19 (1998).

2. D. L Meier, S. Koide, Y. Uchida, Magnetohydrodynamic production of relativistic jets, *Science* **291**, 84-92 (2001).

3. R.D. Blandford, R. L. Znajek, Electromagnetic extraction of energy from Kerr black holes, *MNRAS* **179**, 433-456 (1977).

4. R. D. Blandford, D. G. Payne, Hydromagnetic flows from accretion discs and the production of radio jets, *MNRAS* **199**, 883-903 (1982).

5. J. –P. De Villiers, J. F. Hawley, J. H. Krolik, S. Hirose, Magnetically driven accretion in the Kerr metric. III. unbound outflows, *Astrophys. J.* **620**, 878-888 (2005).

6. J. McKinney, General relativistic magnetohydrodynamic simulations of the jet formation and large-scale propagation from black hole accretion systems, *MNRAS* **368**, 1561-1582 (2006).

7. D. Garofalo, D. A. Evans, R. M. Sambruna, The evolution of radio-loud active galactic nuclei as a function of black hole spin, *MNRAS* **406**, 975-986 (2010).

8. Tchekhovskoy, A., McKinney, J., Prograde and retrograde black holes: whose jet is more powerful?, *MNRAS in press*, **arXiv:1201.4385**, (2012).

9. D. Clery, Worldwide Telescope Aims to Look Into Milky Way Galaxy's Black Heart, *Science*, **335**, 391 (2012).

10. J. Blakeslee, *et al.*, The ACS Fornax Cluster Survey. V. Measurement and Recalibration of Surface Brightness Fluctuations and a Precise Value of the Fornax-Virgo Relative Distance, *Astrophys. J.* **694**, 556-572 (2009).

11. K. Gebhardt, *et al.*, The black hole mass in M87 from Gemini/NIFS adaptive optics observations, *Astrophys. J.* **729**, 119-131 (2011).

12. J. A. Biretta, F. Zhou, F. N. Owen, Detection of proper motions in the M87 jet, *Astrophys. J.* **447**, 582-596 (1995).

13. E. S. Perlman, J. A. Biretta, W. B. Sparks, F. D. Macchetto, J. P. Leahy, the optical-near-infrared spectrum of the M87 jet from hubble space telescope observations, *Astrophys. J.* **551**, 206-222 (2001).

14. H. L. Marshall, *et al*., A high-resolution X-ray image of the jet in M87, *Astrophys. J.* **564**, 683-687 (2002).

15. W. Junor, J. A. Biretta, M. Livio, Formation of the radio jet in M87 at 100 Schwarzschild radii from the central black hole, *Nature* **401**, 891-892 (1999).

16. T. P. Krichbaum, *et al.,* Sub-milliarcsecond imaging of Sgr A* and M 87, *Journal of Physics Conf. Series* **54**, 328-334 (2006).

17. C. Ly, R. C. Walker, W. Junor, High-frequency VLBI imaging of the jet base of M87, *Astrophys. J.* **660**, 200-205 (2007).

18. Y. Y. Kovalev, M. L. Lister, D. C. Homan, K. I. Kellermann, The inner jet of the radio galaxy M87, *Astrophys. J.* **668**, L27-L30 (2007).

19. K. Asada, M. Nakamura, The structure of the M87 jet: a transition from parabolic to conical streamlines, *Astrophys. J.* **745**, L28 (2012).

20. See Supplementary Material.

21. J. C. McKinney, R. D. Blandford, Stability of relativistic jets from rotating, accreting black holes via fully three-dimensional magnetohydrodynamic simulations, *MNRAS* **394**, L126-L130 (2009).

22. J. Gracia, N. Vlahakis, I. Agudo, K. Tsinganos, S. V. Bogovalov, synthetic synchrotron emission maps from MHD models for the jet of M87, *Astrophys. J.* **695**, 503-510 (2009).

23. K. Hada, *et al*., An origin of the radio jet in M87 at the location of the central black hole, *Nature* **477**, 185-187 (2011).

24. A. P. Marscher, *et al.*, The inner jet of an active galactic nucleus as revealed by a radio-to-$\gamma$-ray outburst, *Nature* **452**, 966-969 (2008).

25. A. E. Broderick, A. Loeb, Imaging the black hole silhouette of M87: implications for jet formation and Black Hole spin, *Astrophys. J.* **697**, 1164-1179 (2009).

26. P. E. Hardee, Stability properties of strongly magnetized spine-sheath relativistic jets, *Astrophys. J.* **664**, 26-46 (2007).

27. C. P. Fragile, Effective inner radius of tilted black hole accretion disks, *Astrophys. J.* **706**, L246-L250 (2009).

28. A. Broderick, A. Loeb, R. Narayan, The event horizon of Sagittarius A*, *Astrophys. J.* **701**, 1357, (2009).

29. Gammie, C., Shapiro, S. & McKinney, J., Black hole spin evolution, *Astrophys. J.*, **602**, 312-319, (2004).

30. J. Dexter, J. C. McKinney, E. Agol, The size of the jet launching region in M87, *MNRAS* **421**, 1571-1528 (2012).

31. H. Falcke, F. Melia, E. Agol, Viewing the shadow of the black hole at the Galactic center, *Astrophys. J.* **528**, L13-L16 (2000).

32. Johannsen, T. & Psaltis, D., *et al.*, Testing the no-hair theorem with observations in the electromagnetic spectrum. II. Black hole images, *Astrophys. J.*, **718**, 446-454, (2010).

33. Biretta, J.A., Junor, W. & Livio, M., Evidence for initial jet formation by an accretion disk in the radio galaxy M87, *New Astronomy Reviews*, **46**, 239-245, (2002).

34. A. E. Broderick, R. Blandford, Covariant magnetoionic theory – I. Ray propagation, *MNRAS* **342**, 1280-1290 (2003).

35. A. E. Broderick, A. Loeb, Imaging optically-thin hotspots near the black hole horizon of Sgr A* at radio and near-infrared wavelengths, *MNRAS* **367**, 905-916 (2006).

36. V. L. Fish, *et al.*, 1.3 mm wavelength VLBI of Sagittarius A*: detection of time-variable emission on event horizon scales, *Astrophys. J.* **727**, L36 (2011).

37. S. S. Doeleman, *et al.*, Structure of Sagittarius A* at 86 GHZ using VLBI closure quantities, *Astron. J.* **121**, 2610 (2001).

38. S. S. Doeleman, *et al.*, Event-horizon-scale structure in the supermassive black hole candidate at the Galactic Centre, *Nature* **455**, 78 (2008).

39. A. E. E. Rogers, S. S. Doeleman, J. M. Moran, Fringe detection methods for very long baseline arrays, *Astron. J.* **109**, 1391 (1995).

40. J. M. Bardeen, W. H. Press, S. A. Teukolsky, Rotating black holes: locally nonrotating frames, energy extraction, and scalar synchrotron radiation, *Astrophys. J.* **178**, 347 (1972).

**Acknowledgments:** High frequency VLBI work at MIT Haystack Observatory is supported by grants from the National Science Foundation. The Submillimeter Array is a joint project between the Smithsonian Astrophysical Observatory and the Academia Sinica Institute of Astronomy and Astrophysics. The Submillimeter Telescope is operated by the Arizona Radio Observatory (ARO). ARO is partially supported through the NSF University Radio Observatories (URO: AST-1140030) and ATI (AST-0905844) programs. The James Clerk Maxwell Telescope is operated by the Joint Astronomy Centre on behalf of the Science and Technology Facilities Council of the United Kingdom, the Netherlands Organisation for Scientific Research, and the National Research Council of Canada. Funding for ongoing CARMA development and operations is supported by the NSF and the CARMA partner universities. We thank the NASA Geodesy Program for loan of the CARMA Hydrogen Maser; J. Test, P. Yamaguchi, G. Reiland, J. Hoge and M. Hodges for technical assistance; the staff at all participating facilities; J. Gracia and J. McKinney for providing jet simulation data used in this work; and Xilinx, Inc. for equipment donations. Data used in this paper is available in the Supplementary Materials.

**Fig 1**. Measuring the size of the M87 core with 1.3~mm VLBI. Correlated flux density data from three consecutive days of observing are plotted as a function of baseline length. The CARMA-SMT baselines are shown in red; the two baselines from CARMA dishes to the JCMT are shown in magenta and teal; the SMT-JCMT baseline is shown in blue. Calibration errors of 5% have been added in quadrature to the $1\sigma$ random errors associated with the incoherent fringe search performed on each baseline. The weighted least squares best-fit circular Gaussian model is shown as a solid line, and has a total flux density of 0.98 Jy and a FWHM size of 40.0 micro arcseconds. A hybrid model (dotted line) combines a circular Gaussian of the same size with a thin ring of diameter 40.0 micro arcseconds, which represents the expected shadow feature due to illumination of the central supermassive black hole from behind by a counter-jet in M87. In the hybrid model, the Gaussian and ring components each contribute one-half the flux density. On VLBI baselines shorter than the null spacing, the VLBI interferometric phase is zero, but on baselines beyond the null spacing, the phase is 180 degrees.

**Fig. 2**. Width profile of the M87 jet as a function of distance from the core. Measurements of jet opening angle from the literature (33) were converted to projected jet width, and fit with a power law (dashed black line) of the form $\theta \propto r^\beta$ where $\theta$ is jet width and r is separation from the core. The best fit $\theta \propto r^{0.69}$ indicates a parabola-like profile for the jet with collimation increasing at larger distances as expected from MHD theory, and is in agreement with recent detailed measurements of the M87 jet profile (19). The dark blue horizontal line indicates the size reported in this paper, with the light blue band corresponding to $1\sigma$ uncertainties. To estimate the position relative to the central black hole, we extrapolated the power law fit, which intersects the 1.3mm VLBI size at an apparent core distance of ~1 $R_{SCH}$. GRMHD models, tailored to simulate jet emission from the black hole out to distances of ~100 $R_{SCH}$, are shown as the solid magenta and green lines for jet axis angles to our line of sight of 15 and 25 degrees respectively (21). These close-in simulations also intersect the 1.3mm VLBI size between 1 and 2 $R_{SCH}$ from the black hole. For comparison and illustration, jet width values derived from a separate GRMHD simulation of the larger scale M87 jet (22) are shown as a solid black line for distances larger than ~100 $R_{SCH}$. The red horizontal line indicates the apparent size of the Innermost Stable Circular Orbit (ISCO) for a non-spinning black hole when strong gravitational lensing effects near the black hole are properly accounted for (20).

**Fig. 3**. Diameter of the Innermost Stable Circular Orbit (ISCO) for a black hole of arbitrary spin. The apparent diameter of the ISCO due to the strong gravitational lensing near the black hole was computed using ray-tracing algorithms through Kerr spacetime (34,35). Two scenarios are shown. The black curve is the apparent diameter of an opaque sphere whose radius coincides with the ISCO, and is viewed along the spin axis of the black hole. This distribution approximates a thick accretion disk, which is appropriate for M87. The red curve is the apparent diameter for a ring of emission at the ISCO as viewed in the orbital plane (analytic expressions given in (20). Solid lines show prograde ($a > 0$) orbits and dashed lines show retrograde ($a < 0$) orbits. The 1.3mm VLBI size derived in this work is shown as a horizontal blue line with a cyan band marking the +/- $1\sigma$ uncertainty. The $3\sigma$ upper limit on the 1.3mm VLBI size corresponds to a lower limit on the black hole spin of ($a > 0.2$).

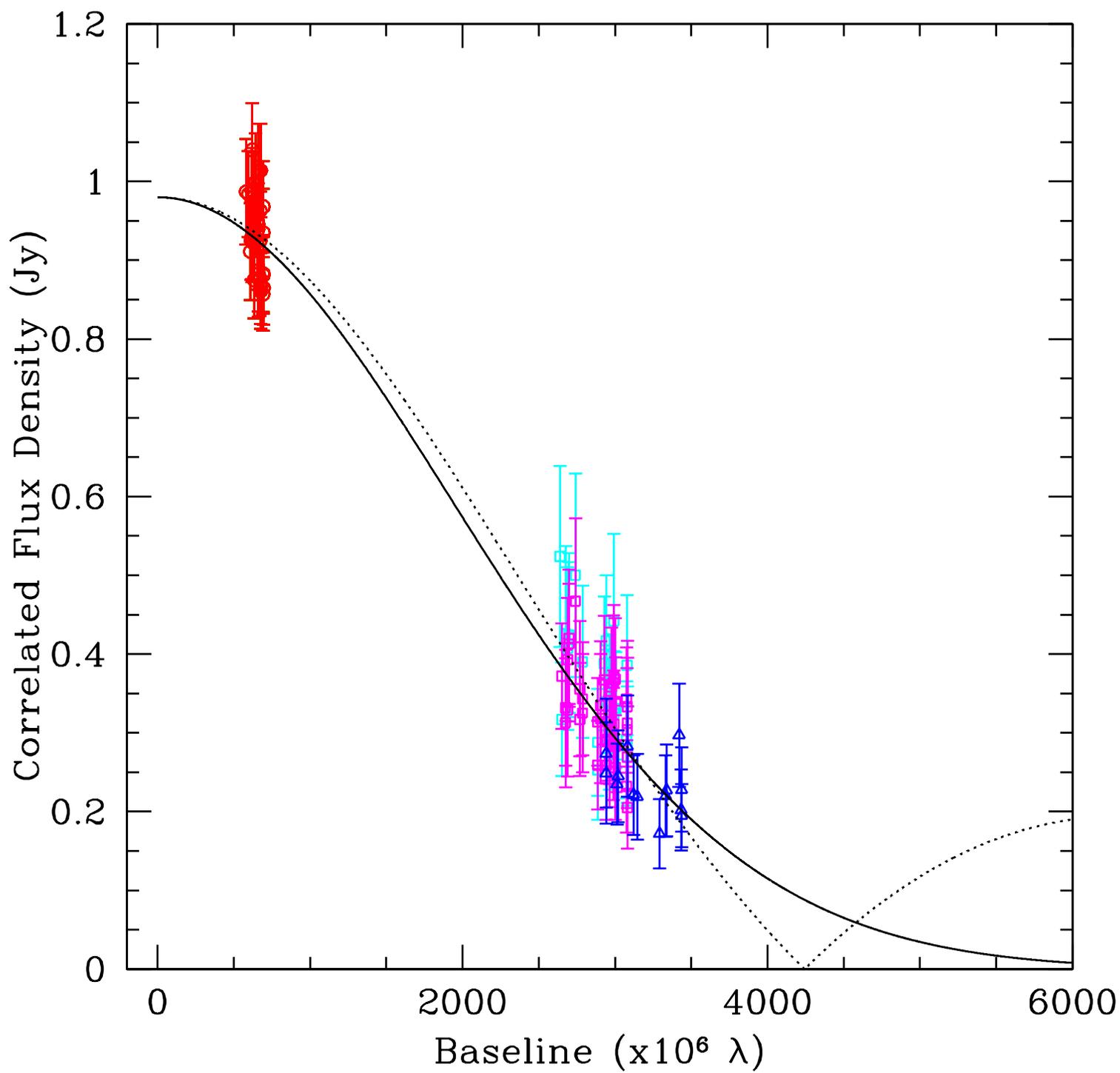

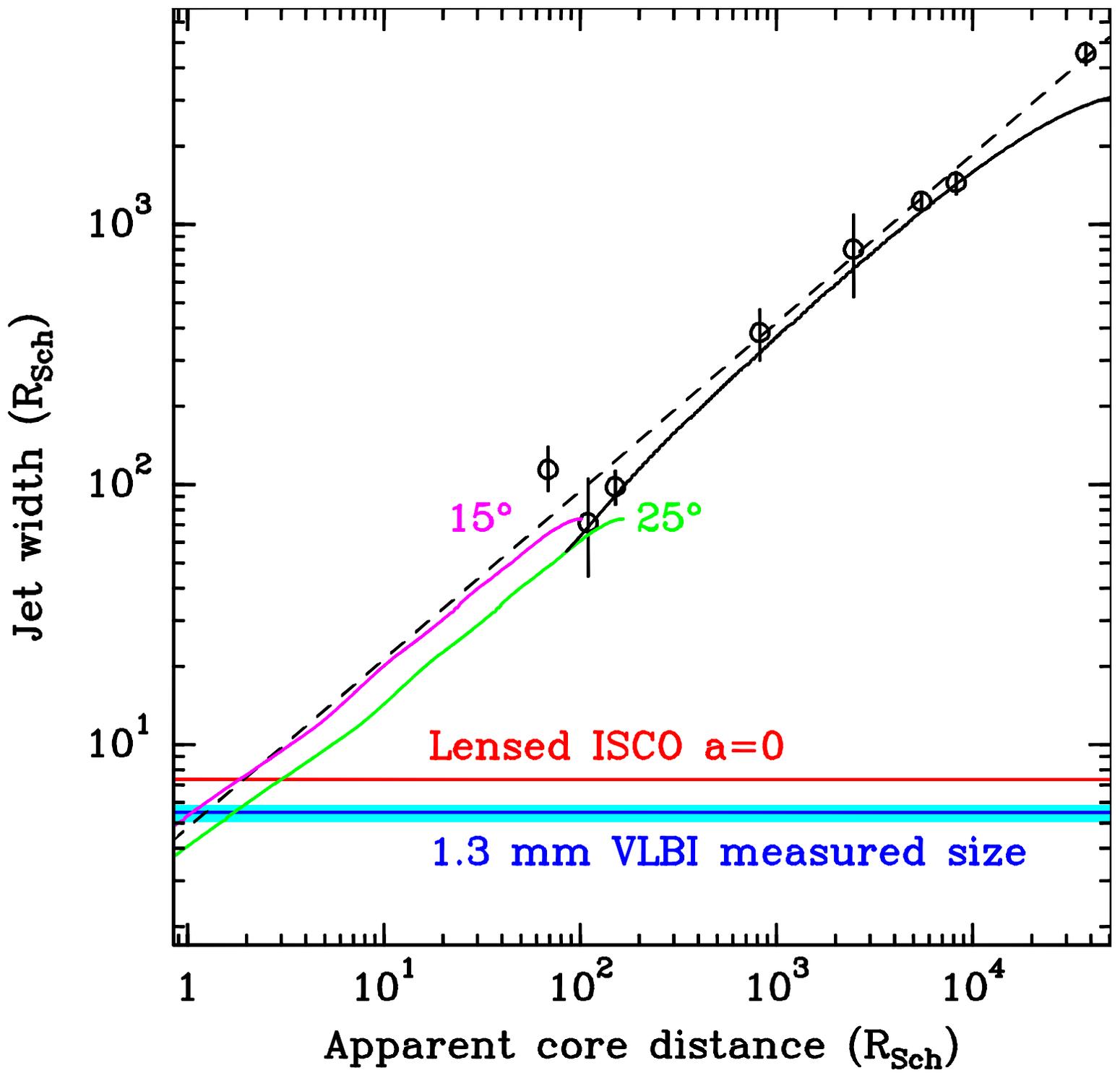

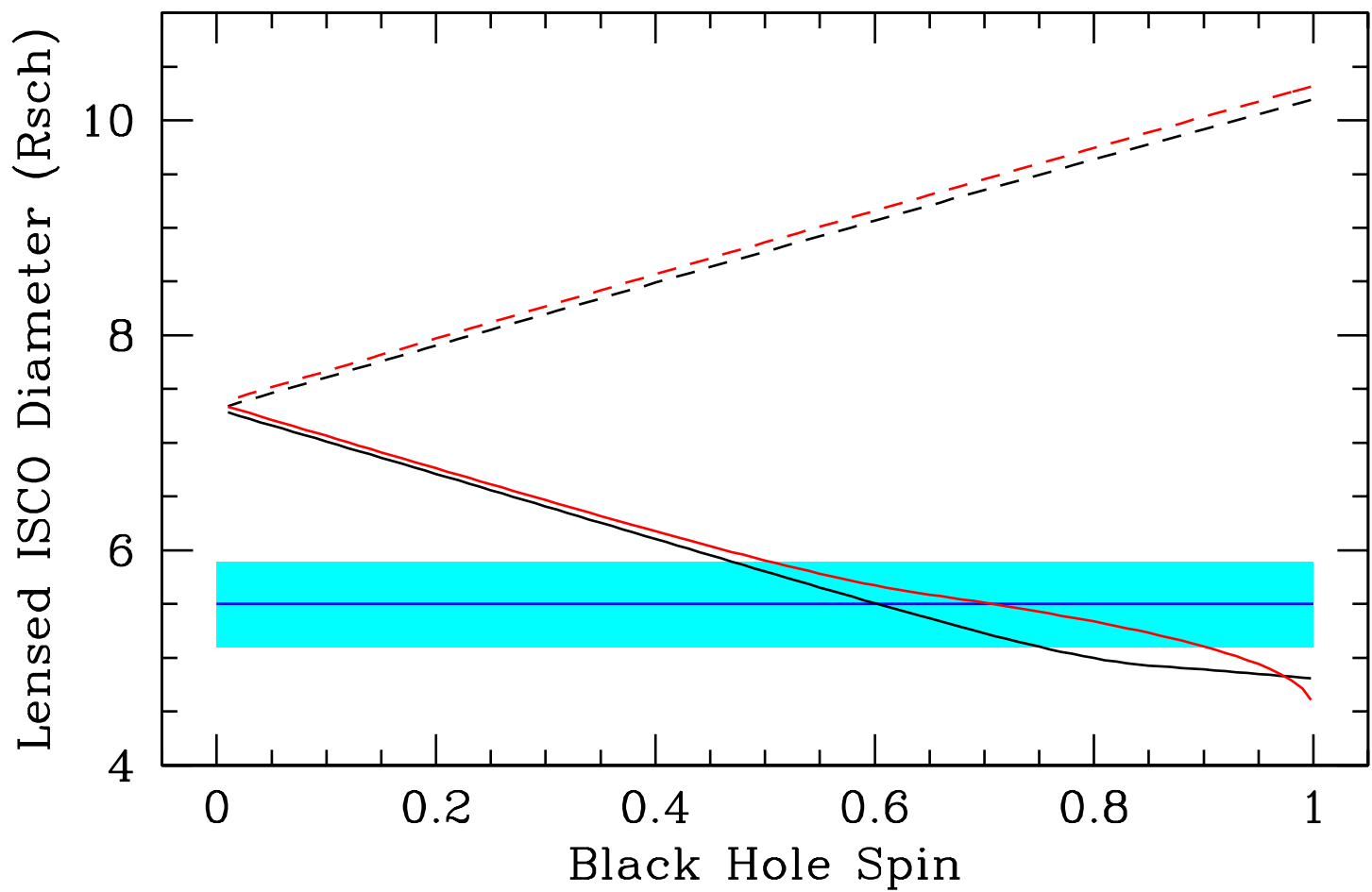